%% file: eprint.tex
\documentclass[12pt]{article}
\usepackage{graphicx}

\def\Title#1{\begin{center} {\Large #1 } \end{center}}
\def\Author#1{\begin{center}{ \sc #1} \end{center}}
\def\Address#1{\begin{center}{ \it #1} \end{center}}

\newenvironment{Abstract}{\begin{quotation}  }{\end{quotation}}
\newenvironment{Presented}{\begin{quotation} \begin{center} 
             PRESENTED AT\end{center}\bigskip 
      \begin{center}\begin{large}}{\end{large}\end{center} \end{quotation}}

\input econfmacros.tex

\begin{document}
\begin{titlepage}

\vfill
\Title{Underlying Event Measurements at CMS}
\vfill
\Author{Rajat Gupta}
\Address{Panjab University, Chandigarh}
\vfill
\begin{Abstract}
Measurements of Underlying Event activity using proton-proton collision data collected by the CMS detector will be presented. To check the energy dependence of the underlying event activity, results are compared with previous measurements from different experiments at different centre-of-mass energies.\end{Abstract}
\vfill
\begin{Presented}
Presented at EDS Blois 2017, Prague, \\ Czech Republic, June 26-30, 2017
\end{Presented}
\vfill
\end{titlepage}
\def\thefootnote{\fnsymbol{footnote}}
\setcounter{footnote}{0}

\section{Introduction}

The combination of particle production from multiple parton interactions (MPI) (excluding parton-parton scattering with the highest momentum transfer) and beam-beam remnant (BBR) interactions is commonly called the underlying event (UE). The UE usually produces particles at low transverse momenta $p_{T}$ that cannot be experimentally distinguished from particles produced from initial (ISR) and final state radiation (FSR). These processes cannot be completely described by perturbative quantum chromodynamics (QCD) calculations, and require phenomenological models, whose parameters are tuned by means of fits to data. The properties of the UE are measured as a function of conventional observables related to the impact parameter of the pp collision, such as the average number of charged particles and the scalar sum of their $p_{T}.$ The data are corrected for detector effects using the iterative D'Agostini method~\cite{unfold} and compared to Monte Carlo (MC) event generators, as well as with earlier results at $\sqrt{s}$~=~1.96~\cite{uecdf} and 7 TeV~\cite{uedycms}. In this paper, we present recent results on measurements of UE activity using events with a leading track and leading jet at a centre-of-mass energy of 13 TeV~\cite{CMS-PAS-FSQ-15-007}, and measurements of the UE activity using events with a  Z boson (with muonic decay) at a centre-of-mass energy of 13 TeV~\cite{CMS-PAS-FSQ-16-008}. The UE measurement with Z boson events is complementary to measurements with a leading jet and leading track, which corroborate the universality of the UE.

\section{UE measurement using leading charged particle tracks and charged jets}

The UE measuerement is performed using leading charged particles as well as leading charged particles jets as reference objects at the centre-of-mass energy of 13 TeV~\cite{CMS-PAS-FSQ-15-007}. A leading charged particle or charged particle jet is required to be produced in the central pseudorapidity region ($|\eta| <$ 2) and with transverse momentum $p_{T} >$ 0.5 ($p_{T}^{jet} >$ 1) GeV for the leading charged particle (charged particle
jet).

The data used in this analysis are selected from an unbiased sample of events whenever there
is a beam crossing in the CMS detector. This corresponds to an integrated luminosity of 281 nb$^{-1}$.

The UE activity is quantified in terms of the average number of charged particles in an event per unit $\Delta\eta\Delta\phi$ area (particle density) and the scalar sum of their $p_{T}$ in an event per unit $\Delta\eta\Delta\phi$ area ($\Sigma p_{T}$ density), with $|\eta|<$ 2 and $p_{T} >$ 0.5 GeV  in the region orthogonal to the azimuthal direction of the leading charged particle or jet, referred to as the transverse region ($60^{\circ} <|\Delta\phi|<$ 120$^{\circ}$). The transverse region can be split into 2 halves depending on the sign of
$|\Delta\phi|$.  The transMAX (transMIN) densities are then defined as the densities in the transverse half with a higher (lower)
activity.  The transDIF density is then defined as the difference of transMAX and transMIN densities. 
Adding all sources of uncertainty in quadrature results in a total systematic uncertainty of about 8--9\%

\begin{figure}[htbp]
\begin{minipage}[t]{.45\textwidth}
\centering
\includegraphics[width=1\textwidth]{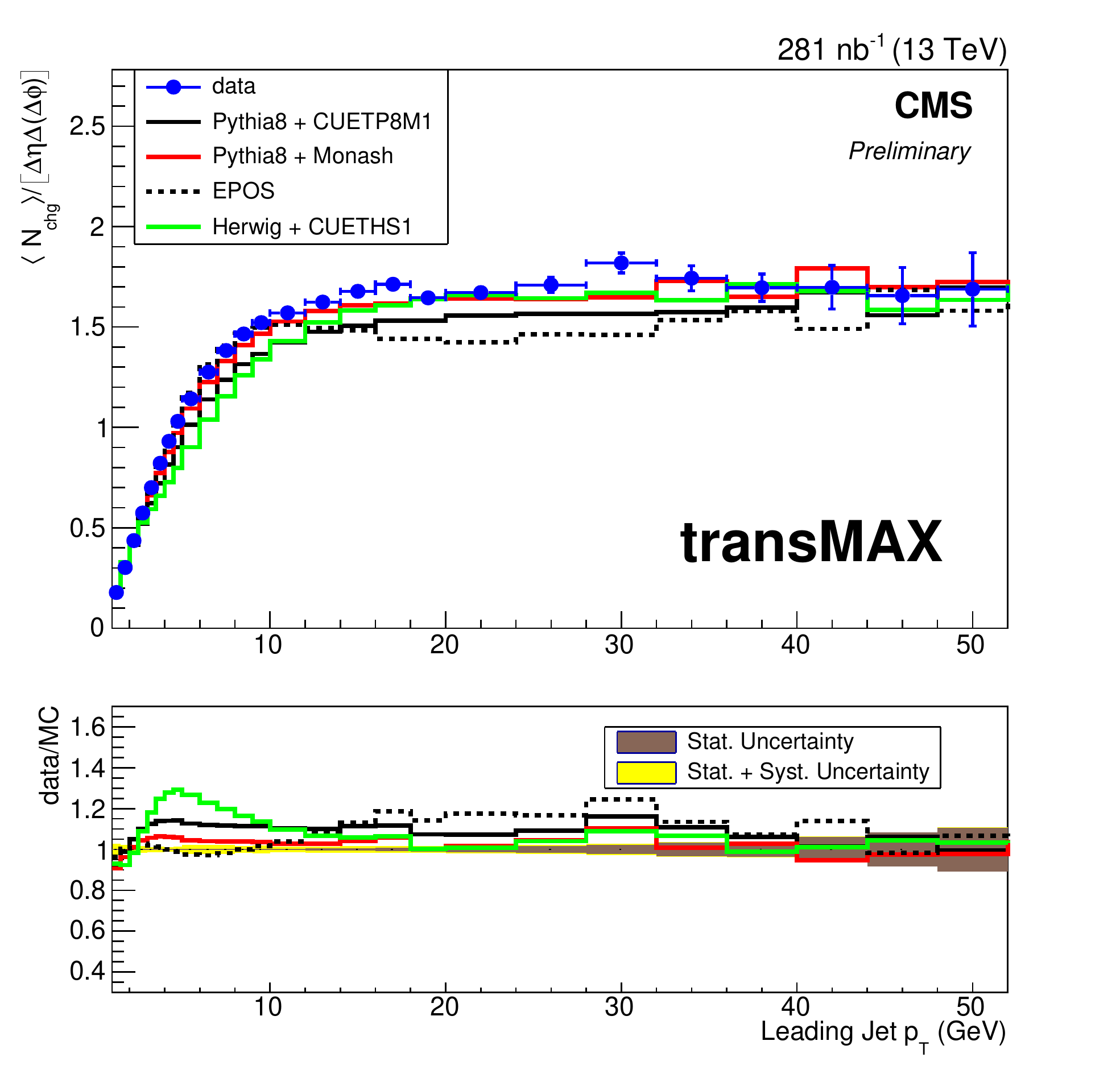}
\caption {{Comparisons of corrected transMAX average particle densities with the various simulations as a function of $p_{T}^{jet}$~\cite{CMS-PAS-FSQ-15-007}.}} \label{fig:TransMAX}
\end{minipage}
\hfill
\begin{minipage}[t]{.45\textwidth}
\centering
\includegraphics[width=1\textwidth]{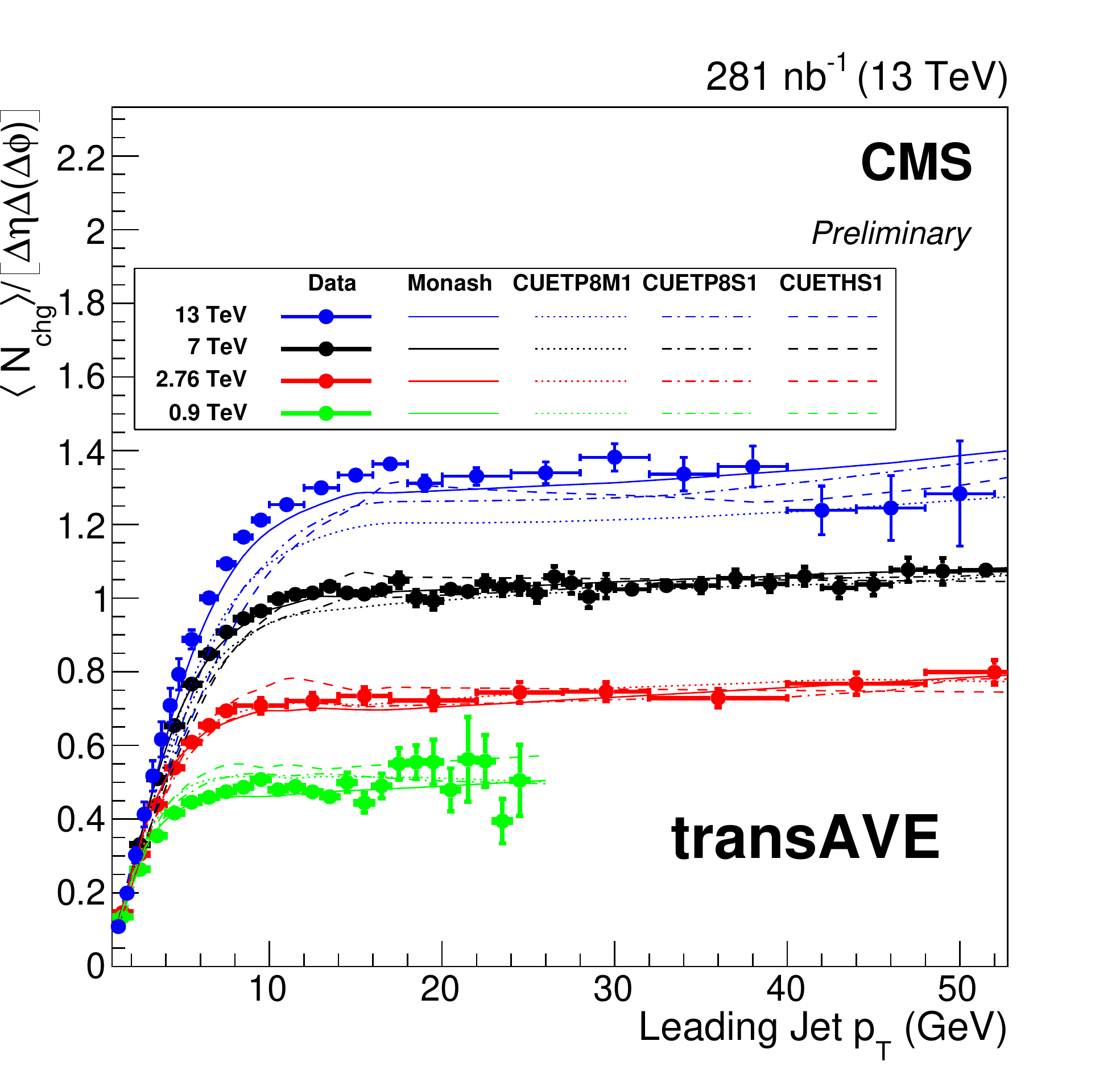}
\caption {{Comparisons of corrected transAVE particle densities with various simulations at $\sqrt{s} =$ 0.9, 2.76, 7, and 13 TeV as a function of $p_{T}^{jet}$~\cite{CMS-PAS-FSQ-15-007}.}} \label{fig:allEnergyComp}
\end{minipage}
\end{figure}
   
The transMAX particle densities are shown in figure~\ref{fig:TransMAX}, as a function of $p_{T}^{jet}$.  The measurements are better described by the Monash tune of PYTHIA8. The PYTHIA8 CUETP8M1 tune describes the measurements within 10$-$20\%. The predictions by the CUETHS1 tune of HERWIG fails in the low $p_{T}$ region. EPOS describes the low $p_{T}$ rising region well but fails to describe the plateau region by 20\%. Distributions for the average
$\Sigma p_{T}$ densities (not shown) also reveal similar behaviour. In all plots, the densities increase sharply up to 5 (12$-$15) GeV and then rise slowly with increasing $p_{T}$ ($p_{T}^{jet}$). The level of agreement between simulations and the measurements falls within 10$-$20\%
in  the  plateau  region  but  differs  in  the  low $p_{T}$ region.   The  sharp  rise is  interpreted in the MC models as due to an increase in the MPI contribution which reaches a plateau at high $p_{T}$.  Comparisons between various MC simulated samples and data across centre-of-mass energies of 0.9, 2.76, 7, and 13 TeV are made for transAVE as a function of  $p_{T}^{jet}$ as shown in figure~\ref{fig:allEnergyComp}. There is a strong rise in the UE activity as a function of the centre-of-mass energy, as predicted by the MC tunes. The  transMIN  (not  shown)  densities  exhibit  a  stronger $\sqrt{s}$ dependence  than  the transDIF (not shown) density, indicating that the activity coming from MPI grows more with $\sqrt{s}$ than that from ISR and FSR.


\section{UE measurement using the Z-boson  process}

A measurement of the UE activity using events with a  Z-boson (with muonic decay) at a 
centre-of-mass energy of 13 TeV~\cite{CMS-PAS-FSQ-16-008} is presented. The Z-boson production is experimentally clean and theoretically well understood, allowing a clear identification of the UE activity. 


The analysis is performed with a data sample of pp collisions corresponding to an integrated luminosity of 2.1 fb$^{\rm -1}$ 
at a centre-of-mass energy of 13 TeV, collected using the CMS detector at the LHC~\cite{Chatrchyan:2008aa}. Events are triggered with the
requirement of at least two isolated muon candidates with $p_{T} >$ 17 GeV and 8 GeV for leading and subleading muons 
respectively.

The offline selection criteria require each event to have at least one well reconstructed primary vertex. 
Both muons are required to lie within a range of $|\eta| <$ 2.4.  The events with two oppositely charged muons are further required to have an invariant mass
(M$_{\mu\mu}$) in the window of 81--101 GeV. After all the selections, there are about 
1.3 million Z candidate events in the data, which is in agreement with simulated samples within 1--2\%. Selected events 
have background contributions, mainly from top-quark and diboson processes, of about 0.3\%.

In the selected Z-boson events, all the tracks with $p_{T}>~$0.5 GeV and $|\eta|<$ 2 are considered for the UE measurements.
The track selection efficiencies for data and simulated samples agree within 4--5\%. 

The UE activity is again quantified in terms of the particle density and their $\Sigma p_{T}$ density. These observables are calculated in different phase-space regions defined with respect to the resultant azimuthal direction of the two selected muons, classified as {\it towards}
region ($|\Delta\phi|<$ 60$^{\circ}$), {\it transverse} region ($60^{\circ} <|\Delta\phi|<$ 120$^{\circ}$), and {\it away} region
($|\Delta\phi|>$ 120$^{\circ}$). 

For the comparison with predictions from different simulations and tuning of model parameters, the UE 
distributions are corrected to the stable charged particle level using the iterative D'Agostini method~\cite{unfold}, which properly considers the bin-to-bin migrations. The unfolded measured distributions can get biased due to the selection criteria and simulated samples used for the 
unfolding. The total systematic uncertainty in the particle and $\Sigma p_{T}$ densities is about 4.8--7.8\%.

The unfolded distributions of the UE activity as a function of $p_{T}^{\mu\mu}$ are compared with predictions from various 
simulations. In order to understand the MPI evolution with centre-of-mass energies, measurements are also compared 
with previous results from Tevatron and LHC.

\begin{figure}[htbp]
\begin{minipage}[t]{.45\textwidth}
\centering
\includegraphics[width=1\textwidth, height=7.7cm]{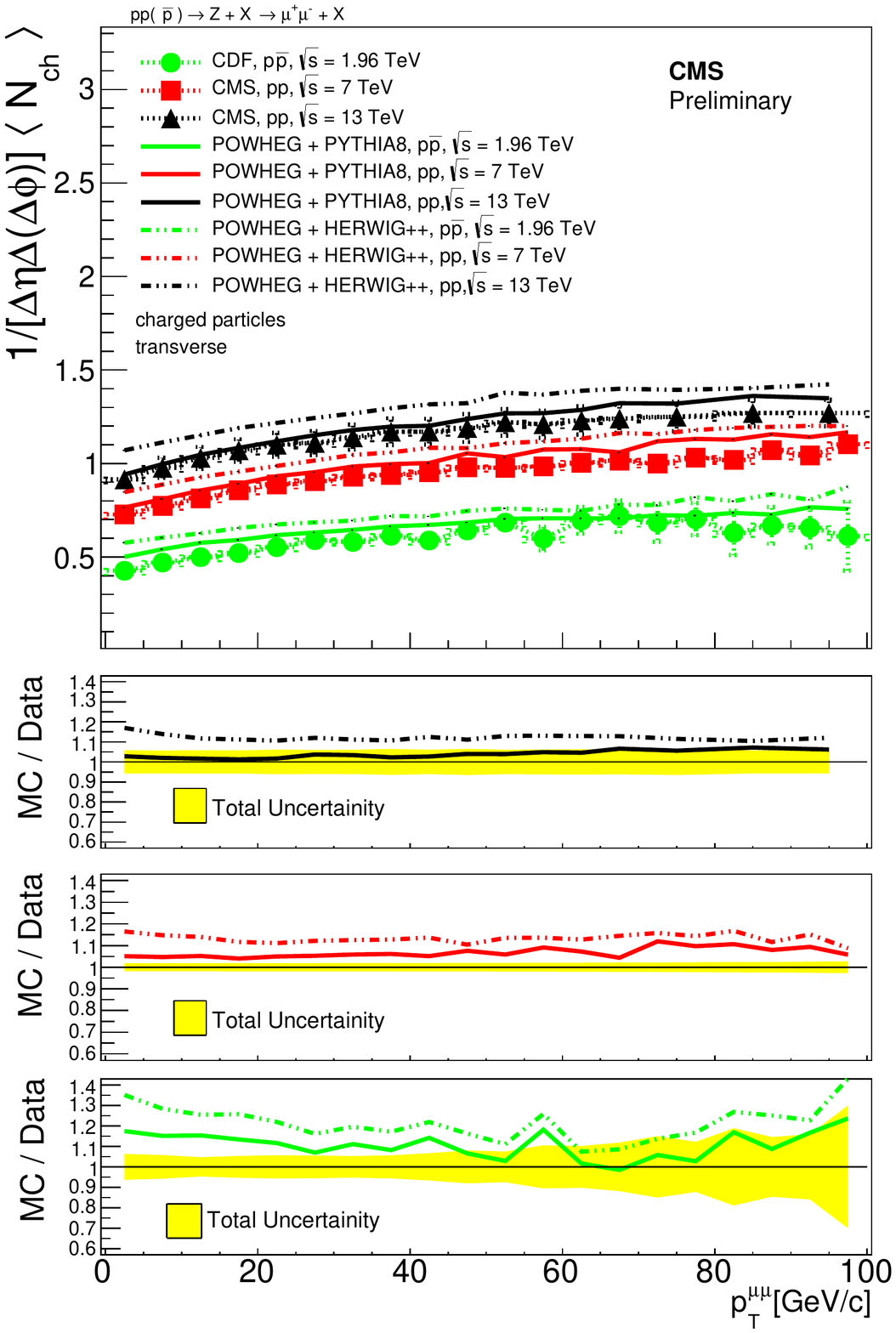}
\caption {{Comparison of UE activity measured at $\sqrt{s}=$13 TeV with the measurements at 7 TeV and 1.96 TeV, by the CMS and CDF 
experiments, for particle density in the {\it transverse} region as a function of $p_{T}^{\mu\mu}$~\cite{CMS-PAS-FSQ-16-008}.}} \label{fig:allEnergyComp2}
\end{minipage}
\hfill
\begin{minipage}[t]{.45\textwidth}
\centering
\includegraphics[width=1\textwidth]{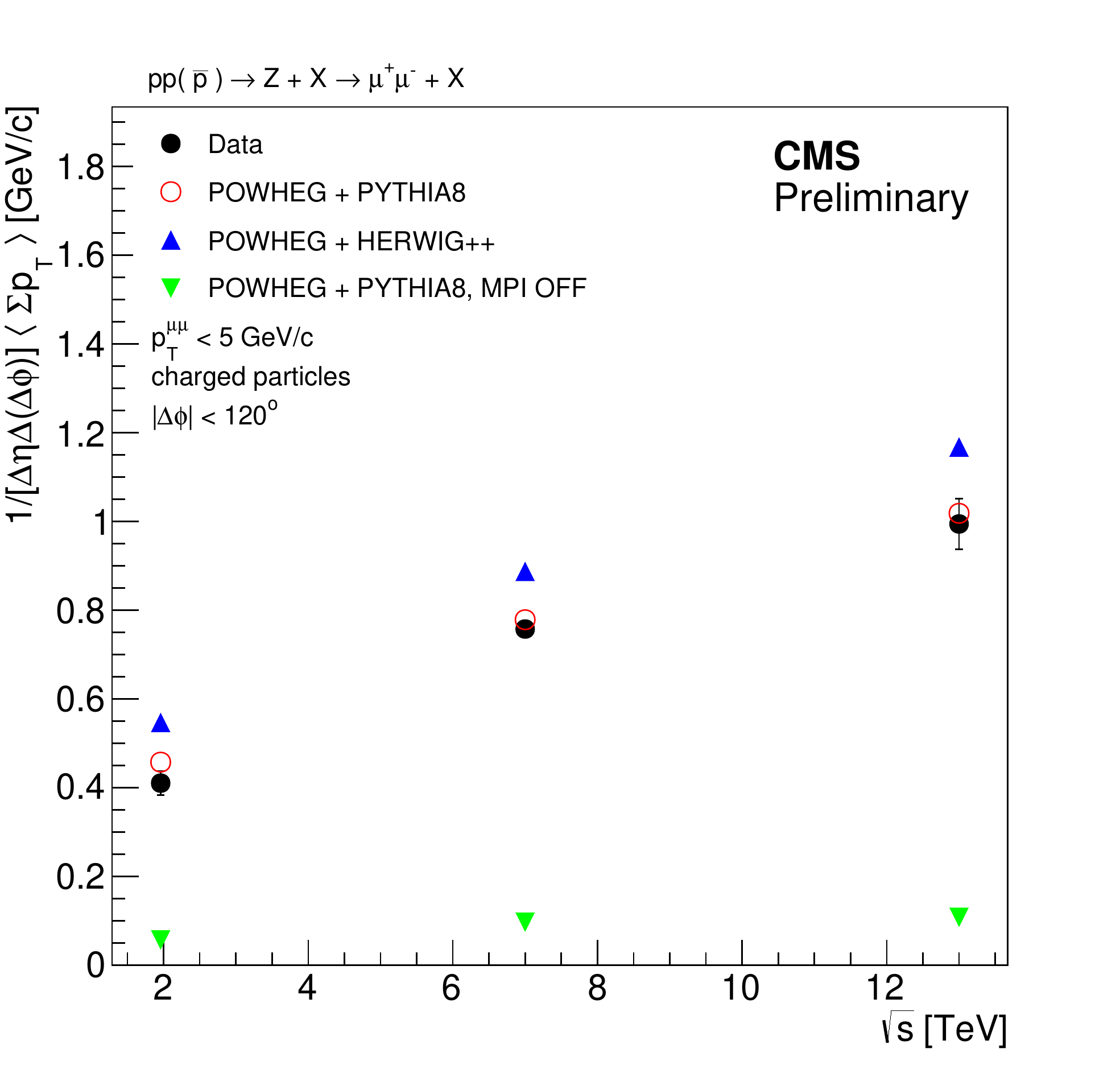}
\caption {{$\Sigma p_{T}$ density, with $p_{T}^{\mu\mu}$~$<$~5~GeV as a function of centre-of-mass 
energy for data, and predictions from simulations by \textsc{powheg} + \textsc{pythia}8, and \textsc{powheg} + \textsc{herwig}++~\cite{CMS-PAS-FSQ-16-008}.}} \label{fig:dataMCpickPt}
\end{minipage}
\end{figure}

Figure~\ref{fig:allEnergyComp2} shows the UE activity as a function of $p_{T}^{\mu\mu}$ at centre-of-mass energies of 
1.96, 7, and 13 TeV. The predictions of the \textsc{powheg} event generator with \textsc{pythia}8, as well as with 
\textsc{herwig}++, are also shown. The ratios of simulations to the measurements are shown in the bottom panel of each plot. 
The \textsc{powheg} + \textsc{pythia}8 predictions describe the measurements within 10\% at centre-of mass energies of 
1.96 TeV and 7 TeV, and within 5\% at 13 TeV. The combination of \textsc{powheg} and \textsc{herwig}++ describes the
measurements within 10--15\%, 10--20\%, and 20--40\% at a centre-of-mass energy of 13 TeV, 7 TeV, and 1.96 TeV respectively.

To quantify the energy dependence of the UE activity, events with a $p_{T}^{\mu\mu}$ smaller than 5 GeV are considered.
An upper cut on $p_{T}^{\mu\mu}$ reduces the radiation contribution and the resulting UE activity comes mainly from MPI. With a 
requirement of $p_{T}^{\mu\mu}$~$<$~5~GeV, the UE activity is similar in the {\it towards} and {\it transverse} regions.
Therefore, the UE activity is combined in these two regions. Figure~\ref{fig:dataMCpickPt} shows the UE activity, after 
an upper cut of 5 GeV on $p_{T}^{\mu\mu}$, as a function of the centre-of-mass energy for data and the simulated samples. The
predictions from \textsc{powheg} + \textsc{pythia}8, without MPI, are also shown. 
It is clear from the comparison of the distributions, with and without MPI, that there is very small contribution from 
radiation, which increases very slowly with centre-of-mass energy. The energy evolution is better described by \textsc{powheg} events hadronized with 
\textsc{pythia}8, whereas hadronization with \textsc{herwig}++ overestimates the UE activity at all energies. 

\section{Summary}

Measurements of underlying event activity at 13 TeV using events with inclusive Z bosons and leading jets/tracks at 13 TeV are presented.
There is scope of further improvements in the underlying event modeling, especially in the energy dependence. The present measurements, in combination with previous results, will be important for further optimization of the model parameters in various simulations.


\end{document}

%% file: econfmacros.tex
\def\beq{\begin{equation}}
\def\eeq#1{\label{#1}\end{equation}}
\def\eeqn{\end{equation}}

\def\beqa{\begin{eqnarray}}
\def\eeqa#1{\label{#1}\end{eqnarray}}
\def\eeqan{\end{eqnarray}}

\let\bar=\overbar

\def\Dslash{\not{\hbox{\kern-4pt $D$}}}
\def\dslash{\not{\hbox{\kern-2pt $\del$}}}

\def\msb{{\bar{\ssstyle M \kern -1pt S}}}

